\documentclass[12pt]{article}
\usepackage{hyperref}
\usepackage{amsmath,amssymb}
\usepackage{graphicx}
\usepackage{color}
\definecolor{myblue}{rgb}{0.14,0.11,0.49}
\definecolor{myred}{rgb}{0.74,0.22,0.15}
\definecolor{mygreen}{rgb}{0.05,0.52,0.42}
\definecolor{myyellow}{rgb}{0.96,0.92,0.13}
\definecolor{myorange}{rgb}{1,0.61,0.36}
\definecolor{mypurple}{rgb}{0.71,0.02,1}
\definecolor{noir}{gray}{0.} 

\newcommand{\Couleur}[1]{\textcolor{noir}{#1}}

\definecolor{htc}{rgb}{1,1,1} 

\def\be{\begin{equation}}
\def\ee{\end{equation}}
\def\bea{\begin{eqnarray}}
\def\eea{\end{eqnarray}}
\def\bc{\begin{center}}
\def\ec{\end{center}}
\def\bi{\begin{itemize}}
\def\ei{\end{itemize}}
\def\bs{\begin{slide}}
\def\es{\end{slide}}
\def\dd{\mathrm{d}}

\def\noi{\noindent}
\title{{\bf Is spacetime as physical as is space?}}
\author{
Mayeul Arminjon\\
\small\it Univ. Grenoble Alpes, CNRS, Grenoble INP, 3SR, F-38000 Grenoble, France
} 
\date{}
			     
\begin{document}

\maketitle
\begin{abstract}
\noi Two questions are investigated by looking successively at classical mechanics, special relativity, and relativistic gravity: first, how is space related with spacetime? The proposed answer is that each given reference f\mbox{}luid, that is a congruence of reference trajectories, defines a physical space. The points of that space are formally defined to be the world lines of the congruence. That space can be endowed with a natural structure of 3-D differentiable manifold, thus giving rise to a simple notion of spatial tensor --- namely, a tensor on the space manifold. The second question is: does the geometric structure of the spacetime determine the physics, in particular, does it determine its relativistic or preferred-frame character? We find that it does not, for different physics (either relativistic or not) may be defined on the same spacetime structure --- and also, the same physics can be implemented on different spacetime structures.\\
\noi \textsl{MSC}: 70A05 [Mechanics of particles and systems: Axiomatics, foundations]\\
 70B05 [Mechanics of particles and systems: Kinematics of a particle]\\
 83A05 [Relativity and gravitational theory: Special relativity]\\
 83D05 [Relativity and gravitational theory: Relativistic gravitational theories other than Einstein's]\\
 \textsl{Keywords}: Affine space; classical mechanics; special relativity; relativistic gravity; reference fluid.

\end{abstract}



\section{Introduction and Summary}\label{Intro}

What is more ``physical": space or spacetime? Today, many physicists would vote for the second one. Whereas, still in 1905, spacetime was not a known concept. It has been since realized that, even in classical mechanics, space and time are coupled: already in the minimum sense that space does not exist without time and vice-versa, and also through the Galileo transformation. An axiomatic scheme of our world --- whether a relativistic or a non-relativistic scheme --- is easier to build by starting from a (model) spacetime. In such a model, time and space exist at once, albeit in a mixed state. This asks the questions of how to define time and space from spacetime. The proper time on a trajectory, at least, is well defined in a Lorentzian spacetime. How to define space from spacetime is less obvious and is one of the two problems that are discussed in this paper. This question has not been often discussed before. In the pre-relativistic concept, there simply was no consideration of spacetime. In today's relativistic view, space either (often) is more or less explicitly considered as an obsolete concept or (sometimes) is assumed to be essentially a space-like hypersurface in spacetime (see e.g. Lachi\`eze-Rey \cite{Lachièze-Rey2003}). The present author argued that space is definitely needed in relativistic physics as well, and that the points of space have to exist at least for some interval of time, hence cannot be single events, and instead have to be defined as world lines in spacetime \cite{A44, A52}. The starting point for that approach was the notion of system of reference as it has been used, in particular, by Landau \& Lifshitz \cite{L&L}, by M\o ller \cite{Moller1952}, and by Cattaneo \cite{Cattaneo1958}. Essentially, all of these authors considered that a system of reference can be defined by the data of a coordinate system $(x^\mu)\ (\mu=0,...,3)$ on the spacetime manifold $\mathrm{V}$, and that the ``spatial position" in that system of reference is defined by the three ``spatial" coordinates $x^i\ (i=1,2,3)$, so that the system of reference is unchanged if one changes the coordinates in such a way that the change of the spatial coordinates $x^i$ does not depend on the time coordinate $x^0$; they defined a ``spatial metric" $h_{ij}$ whose indices are only spatial, and which allows one to define spatial distances (at least in an infinitesimal neighborhood of a given point of spacetime). In particular, Cattaneo \cite{Cattaneo1958} defined an admissible ``{\it f\mbox{}luid of reference}" made of ``{\it reference particles}" having time-like world lines. (Inside quotations, all italics  are in the original texts.) He started from an admissible coordinate system $(x^\mu)$ on the spacetime manifold $\mathrm{V}$, i.e., such that the condition
\be\label{admissible-coordinates}
g_{00}>0, \qquad g_{ij}\dd x^i \dd x^j <0   \quad (i,j=1,2,3)
\ee
be verified in these coordinates, where ${\boldsymbol{g}}$ is the (Lorentzian) spacetime metric on $\mathrm{V}$. He observed that this condition ``permits the co-ordinate lines $x^0 = \mathrm{var.}$ to be interpreted as time tracks of $\infty ^3$ ideal particles which in their totality constitute the {\it physical system of
reference S} associated to the chosen system of co-ordinates" \cite{Cattaneo1958}. However, none of these authors, even Cattaneo, considered that system of reference as ``the space", hence they even less considered a structure of differentiable manifold on it: in fact it is not clear that they envisaged the physical congruence of reference world lines as a set in the mathematical sense. A progress in that direction was made (for the case of a rotating frame of reference) by  Wahlquist and Estabrook \cite{WahlquistEstabrook1966}: they considered ``the `quotient space' --- the 3-dimensional manifold of comoving coordinates, $\chi^a$ " and added: ``Geometrically, we may picture the quotient space as a reduction of space-time obtained when all events lying on each world line of the timelike congruence are identified." See also Rizzi and Ruggiero \cite{RizziRuggiero2002}. 
As we found recently, it turns out that the general formulation was initiated by Norton \cite{Norton1985} who named ``frame of reference" precisely what Cattaneo called a system of reference, but added: ``we formally define the relative space $R_F$ of a frame of reference $F$ in a four-dimensional manifold $M$ as follows. $F$ defines an equivalence relation $f$ under which points $p$ and $p'$ of $M$ are equivalent if and only if they lie on the same curve $c$ of $F$. The relative space $R_F$ is the quotient manifold $M/f$ and has the curves of $F$ as elements. Coordinate charts of $R_F$ are inherited directly from the coordinate charts of $M$ which are adapted to the frame, ensuring that $R_F$ has a well-defined local topology. That is, if ${x^i}\ (i= 1,2,3,4)$ is a chart in a neighborhood of $M$ adapted to $F$, then there will be a chart $y^i \ (i= 1,2,3)$ in the corresponding neighborhood of $R_F$ for which $y^i(c)=x^i(p)\ (i= 1,2,3)$ whenever $p$ lies on $c$." This is essentially what we did in detail in Refs. \cite{A44,A52} (without knowing that work of Norton \cite{Norton1985}). However, Norton's quotation above defines in fact a program that leaves many questions, such as: how exactly are defined the curves $c$, including their domain of definition? When  does the quotient set $M/f$ have a Hausdorff topology and when is $M/f$ a differentiable manifold? When does it exist adapted charts? How to define the ``relative space" if we start from one spacetime chart (generally local only), as did Cattaneo, and what is the relation with a ``global" case (also to be defined)? 
etc. We wrote independently \cite{A16} a similar (though less precise) ``program" in the context of any theory of gravitation with a curved spacetime: ``Any possible {\it reference frame} $\mathcal{F}$ [is] physically defined by a {\it spatial network of `observers'} (...). So we have a spacetime manifold $\mathrm{M}^4$. The elements (points) of the spatial network cannot be identified with {\it points} in that manifold but with `world lines', thus with {\it lines} in space-time. Hence, from the point of view of `space-time',  a reference frame is a 3-D differentiable manifold $\mathrm{N}$ whose each point is a (time-like) differentiable mapping from the real line into the space-time $\mathrm{M}^4$. " Only by thus considering explicitly the space associated with a system of reference as a 3-D differentiable manifold $\mathrm{N}$ (whose elements are the world lines of the reference particles) can one give a simple meaning to the ``spatial metric" and to other spatial objects such as the 3-velocity: they are simply tensors (or tensor fields) on the space manifold $\mathrm{N}$ \cite{A16}. (See Subsect. \ref{Space} below.) However, it remained to formulate a precise definition of that manifold, including its topology and atlas,  and to prove that it is indeed a differentiable manifold. That work has been done first in the case of a local reference frame defined from the data of a coordinate system \cite{A44} and then in the case of a global reference f\mbox{}luid defined by a global vector field on the spacetime manifold $\mathrm{V}$ \cite{A52}. It is striking that each of these two precise definitions works independently of any metric on $\mathrm{V}$ (although of course a physical reference frame should preferably be made of time-like particles and the latter notion of course involves a Lorentzian metric). This is perhaps not so surprising insofar as the notion of a differentiable manifold is more general than that of a pseudo-Riemannian metric: the latter needs the former. However, in the literature on relativistic gravitation, it does not happen often that a spacetime manifold be considered without it being endowed with a Lorentzian metric.
\footnote{\
One reason for this is that in general relativity the spacetime manifold indeed is not fixed independently of the metric: ``points of space-time (events) are not individuated apart from their metrical properties" \cite{Stachel1989}.
}
\\

The other problem that is broached in this paper is the relation between the physics and the geometric structure of the spacetime. This is clearly a vast subject and there are many works about it, e.g. Schr\"odinger \cite{Schrodinger1950}, Trautman \cite{Trautman1970, Trautman1973}. In most of those works, a more or less explicit thesis seems to be that physics  in its deep nature is indeed the geometry of spacetime. Ideally, then, the relation between the physics and the geometric structure of the spacetime should be unique. However, although a spacetime endowed with a geometric structure is a clever, beautiful, and very useful mathematical tool in physics, we see an important f\mbox{}lexibility in the correspondence between the geometry of spacetime and the physics. We showed in previous works that on a given curved Lorentzian spacetime, one may define a new dynamics by extending the special-relativistic form of Newton's second law \cite{A16, A15}. This already indicated that the former relation is not unique, since different physics may be defined on the same spacetime structure.\\ 

Thus, previous works showed two things for relativistic theories of gravitation: (i) starting from a spacetime manifold, the physical space depends on the reference frame and can be defined as the set of the world lines of the reference points, which set can be endowed with a natural structure of differentiable manifold without using any metric. (ii) The relation between the physics and the geometric structure of the spacetime is not unique. It was then natural to investigate the latter relation, as well as the relation between space and spacetime, in the simpler theories that are classical mechanics and special relativity. The technically new part in the present paper consists of that investigation (Sects. \ref{ClassicalMech} and \ref{SR}). The other part of the paper (Sect. \ref{Curved}) details Points (i) and (ii) above, based on our previous work: Subsect. \ref{TwoDynamics} reveals Point (ii) from the works \cite{A16,A9}, and Subsect. \ref{Space} summarizes the works \cite{A44,A52} that established Point (i).\\

We shall begin with examining in some detail the situation of classical mechanics: we shall present a thrifty framework for classical kinematics, by taking the spacetime to be the product of two affine spaces. (Of course the introduction of affine spaces in classical mechanics is not new, e.g. Trautman \cite{Trautman1973}; see in particular Arnold \cite{Arnold1976}, Porta Mana \cite{PortaMana2011}. But, to our knowledge, the approach in Sect. \ref{ClassicalMech} is new.) We shall examine in that framework the transition from one coordinate system to another one: this is necessary in order that the notions of velocity and acceleration be properly defined, even already within one given reference frame. The transition from the ``fixed" space to a ``moving" space will also be discussed in that framework. In this course, we shall be led naturally to introduce the essential notion \cite{WahlquistEstabrook1966, RizziRuggiero2002, Norton1985, A16} according to which the points of space are world lines in spacetime and the physical space is the set of those world lines. That set is naturally endowed with a manifold structure --- more particularly a structure of 3-D affine space, in the case of a set of uniformly moving reference points. It is quite obvious that the product spacetime considered here is well adapted to the non-relativistic physics, e.g., to Lorentz's first electromagnetic theory with an ``ether". However, we shall show that that product spacetime is well adapted also to Galilean relativity --- as well, in fact, as is the ``block Galileo Universe" of Arnold \cite{Arnold1976}. This is because the acceleration is invariant in the transition from the ``fixed" space to a ``moving" space. And also because, as we will show in Subsect. \ref{Galileo}, the role played by the ``fixed" space in the product spacetime can be taken as well by any of the uniformly ``moving" spaces. That a product spacetime can describe Galilean relativity should not be a surprise after all, since the notion of separated space and time was at the root of Newton's mechanics, which obeys Galilean relativity. \\

Just like the spacetime of classical mechanics can be defined either as a product of two affine spaces or as the ``block" four-dimensional affine spacetime, it will be shown that the same is true for the Minkowski spacetime of special relativity. It will also be noted that, once the affine-space structure or the vector-space structure of the Minkowski spacetime is given, it can still be endowed with many different Minkowski metrics.\\

Finally, we shall study our two problems in a general spacetime, basing our ref\mbox{}lection on previous results \cite{A44, A52, A16, A9}. First, we shall recall that an alternative dynamics can be defined on a curved Lorentzian spacetime, provided one has the additional structure given by a preferred reference f\mbox{}luid. Second, we shall summarize the two possible definitions of the space --- either a local one \cite{A44} or a global one \cite{A52} --- associated with a reference f\mbox{}luid in a general spacetime.

\section{Classical Mechanics}\label{ClassicalMech}
\subsection{The Newton-Lorentz Universe without or with Galilean Relativity}\label{N-L}
  \subsubsection{The ``Newton-Lorentz Universe"}

Newton's construction of mechanics started from separated space and time. 
\footnote{\
As is well known, Newton's work built upon many essential prior contributions, e.g. by Descartes, Galileo, Kepler, Hooke; moreover, what we now name ``Newtonian mechanics" includes also essential contributions made after Newton by scientists as important as were Euler and Lagrange --- among many others. However, it seems that Newton has been one of the first scientists (in the sense we now give to this word, which does not include Greek philosophers) to propose some axiomatic definition of the physical concepts of space and time.
}
This can be conciled with the spacetime formalism, by considering a product spacetime manifold:
\be\label{time-space-product}
\Couleur{\mathrm{V}_\mathrm{N-L}=\mathrm{A}^1\times \mathrm{A}^3},
\ee
where \Couleur{$\mathrm{A}^1$} is Newton's ``absolute time" axis and where \Couleur{$\mathrm{A}^3$} is Newton's ``absolute space" (a concept that is akin to Lorentz's concept of the ether). Nowadays we may precisely define \Couleur{$\mathrm{A}^1$} as a one-dimensional affine space, and similarly we may define \Couleur{$\mathrm{A}^3$} as a three-dimensional affine space. (Here the superscript just indicates the dimension, defined below; it does not mean a Cartesian power.) An ``affine space" \Couleur{$\mathrm{A}$} is the mathematical structure which underlies the notion of physical space of the high school (when \Couleur{$\mathrm{dim}\,\mathrm{A} =3$}, and except for the metric notions, which are not included in the affine structure): in brief, it is ``a vector space without an origin". Formally (e.g. \cite{Berger1984,CrampinPirani1986,Tarrida2011}): \\

\vspace{2mm}
\noi The set $\mathrm{A}$ is an {\it affine space} iff there is a real vector space \Couleur{$\mathrm{E}$} and a mapping:
\be\label{extern add}
\Couleur{\mathrm{A}\times \mathrm{E} \rightarrow  \mathrm{A}:\quad (a,v)\mapsto b=a+v}
\ee
(that is an action of the additive group of $\mathrm{E}$ on $\mathrm{A}$), such\ that
\bea
(\mathrm{AS1}) & \quad \Couleur{\forall a\in \mathrm{A},\quad a+0_\mathrm{E} = a}. \nonumber \\
(\mathrm{AS2}) & \quad \Couleur{\forall a\in \mathrm{A},\ \forall v,w\in \mathrm{E},\quad (a+v)+w=a+(v+w)}. \nonumber\\
(\mathrm{AS3}) & \quad \Couleur{\forall a\in \mathrm{A},\ T_a:\, \mathrm{E}\rightarrow \mathrm{A},\quad v \mapsto a+v,}\ \mathrm{is\ a\ bijection}.\nonumber
\eea
Axiom (AS3) states that, for any two points $a,b \in \mathrm{A}$, there exists a unique vector $v \in \mathrm{E}$, such that $b = a+v$. One notes $v = b-a$. More explanations about the meaning of this classical definition can be found in the literature, e.g. in Refs. \cite{PortaMana2011, CrampinPirani1986}. The vector space \Couleur{$\mathrm{E}$} is called the {\it translation space} of \Couleur{$\mathrm{A}$}. The dimension of \Couleur{$\mathrm{A}$} is: \ \, \Couleur{$\mathrm{dim}\,\mathrm{A}\equiv \mathrm{dim}\,\mathrm{E}$}. This will be assumed finite.

  \subsubsection{The Newton-Lorentz Universe without Relativity}\label{N-L without R}

With \Couleur{$\mathrm{V}_\mathrm{N-L}=\mathrm{A}^1\times \mathrm{A}^3$}, we have obviously a preferred time: 
\be
\Couleur{T(X) \equiv \mathrm{Pr}_1(X) \in \mathrm{A}^1 \quad [X=(T,x)\in \mathrm{V}_\mathrm{N-L}]},
\ee
and a preferred spatial position: 
\be
\Couleur{x(X) \equiv \mathrm{Pr}_2(X)\in \mathrm{A}^3\quad [X=(T,x)\in \mathrm{V}_\mathrm{N-L}]}.
\ee
This corresponds well with Lorentz's first electromagnetic theory, which predicted effects of a motion through the ``ether" (even if this occurred only from the second-order in $v/c$). The name ``Newton-Lorentz Universe" was introduced by the Romanian theoretician Eugen So\'os \cite{Soos1992} to designate a product of the type ``Time"$\,\times\,$``Space", endowed with a physics which does not necessarily obey the relativity principle. 
\footnote{\
Akin to the N-L Universe (\ref{time-space-product}) is the ``Aristotle spacetime" $A^4 = E^1 \times E^3$, with $E^1$ a 1-D Euclidean space and $E^3$ a 3-D one \cite{Chaverondier2008}. 
}


  \subsubsection{Spatial Points as Trajectories or as World Lines}

The most natural definition of a ``moving point" is as a trajectory in space: $t \mapsto x=g(t)\in \mathrm{A}^3$, parameterized by the real time variable $t$ [which will be related around Eq. (\ref{Change time}) with the ``affine" time variable $T$]. The particular case of a point $x$ that is fixed in the ``absolute" space $ \mathrm{A}^3$ corresponds of course with a ``trajectory at rest", $\psi _x: t \mapsto x=\mathrm{Constant} \in \mathrm{A}^3$. Given a point \Couleur{$x \in \mathrm{A}^3$}, we may also define the {\it world line}
\be\label{l_x}
l_x \equiv \{(T,x);\ T\in \mathrm{A}^1\} = \mathrm{A}^1 \times \{x \}.
\ee
Thus $\ l_x $ is a one-dimensional subset of $ \mathrm{V}_\mathrm{N-L}=\mathrm{A}^1 \times \mathrm{A}^3$. All events \Couleur{$X \in l_x$} have the same spatial position \Couleur{$x \in \mathrm{A}^3$}, and it is equivalent to give one point $x$, or the trajectory $\psi _x$, or still the world line $l_x$. Hence, the position of an event \Couleur{$X =(T,x)\in \mathrm{A}^1 \times \mathrm{A}^3$} in the space $ \mathrm{A}^3$ may also be defined, instead of giving the point $x$, by specifying the constant trajectory $\psi _x$, or still by specifying the world line \Couleur{$l_x$}. Indeed \Couleur{$l_x$} is a representation of the trajectory of one point, in the present case a point that has no motion with respect to the ``preferred reference space". This identification of a spatial point with a trajectory or with a world line may seem artificial in that very simple case, but it becomes mandatory as soon as one considers {\it moving} points. A moving point is naturally defined by a general trajectory, as already said; but almost equivalently, it can be defined as a general world line $\ l \subset \mathrm{A}^1 \times \mathrm{A}^3$, that is, as the {\it image} of a curve in spacetime: $l=C(\mathrm{I})$ where $\mathrm{I}$ is an interval of $\mathbb{R}$ and $C: \xi \mapsto C(\xi )=X \in \mathrm{V}_\mathrm{N-L}$ is the parameterized smooth curve. 
\footnote{\
The parameterization (by $\xi $) of the world line serves to ensure that $l$ is one-dimensional. In contrast with the parameterization of the trajectory, it may be changed very generally without changing the motion of the moving point. Indeed the exact time dependence of its spatial position is contained in the definition of the world line as a one-dimensional subset of the {\it spacetime}. See after Eq. (\ref{x(t)}) below.
}

  \subsubsection{Coordinates, Velocity, and Acceleration in the Affine Spacetime}\label{Coord_aff}

On an affine space $\mathrm{A}^n$, with dimension $n$, there are preferred coordinate systems, which are the {\it affine charts}. Any such chart $\chi  $ is defined by the data of two things: an arbitrary origin point $X_0 \in \mathrm{A}^n$, which maps to the zero vector of the ``arithmetic" $n$-dimensional vector space $\mathbb{R}^n$; and an arbitrary linear bijection $\Lambda$ from the vector space $\mathrm{E}^n$ (the translation space of $\mathrm{A}^n$) onto $\mathbb{R}^n$. Thus:
\be
\chi  :  \mathrm{A}^n \rightarrow \mathbb{R}^n,\quad X \mapsto \Lambda (X-X_0).
\ee 
(Of course the superscript $n$ does mean a Cartesian power for $\mathbb{R}^n$ but not for $\mathrm{A}^n$ or $\mathrm{E}^n$.) In particular, for the Newton-Lorentz universe (\ref{time-space-product}), the charts that are compatible with its product structure have a privileged status: given any $T_0\in \mathrm{A}^1$, any $x_0 \in \mathrm{A}^3$, and any linear bijections $\lambda : \mathrm{E}^1\rightarrow \mathbb{R}$ and $L: \mathrm{E}^3\rightarrow \mathbb{R}^3$, one defines
\be\label{Product chart}
\Phi :(T,x) \mapsto (\theta (T),\phi (x))=(\lambda (T-T_0),\,L(x-x_0))\in \mathbb{R} \times \mathbb{R}^3\simeq \mathbb{R}^4.
\ee
If $\theta :T\mapsto t=\lambda (T-T_0)\ $ is an affine chart of $\mathrm{A}^1$, then any other one, say $\theta '$, has the form
\be\label{Change time}
T \mapsto t'=at+b\qquad (t\equiv \theta (T),\ t'\equiv \theta '(T), \quad a,b\in \mathbb{R},\ a\ne 0),
\ee
as is easy to check. Therefore, we shall fix the chart $\theta $ of $\mathrm{A}^1$. It thus amounts to fixing the origin of time and the time unit (and also the time arrow: a change of the time chart can be said to respect that arrow if $a>0$). \\

Let a moving point be given by a world line $l$ in $\mathrm{V}_\mathrm{N-L}$:
\be\label{World line l}
l = \{C(\xi )\equiv (\hat{T}(\xi ),\hat{x} (\xi ))\in \mathrm{A}^1\times \mathrm{A}^3;\ \xi \in \mathrm{I} \}.
\ee
The time is defined along this world line by
\be\label{time along l}
t=\theta (\hat{T} (\xi ))\equiv f(\xi ),
\ee
and we assume that it is a monotonic function of $\xi$, meaning that $f$ is invertible. This allows us to parameterize the world line $l$ with the time $t$; thus in particular we get the trajectory of the moving point:
\be\label{x(t)}
x = \hat{x} (\xi)=\hat{x}(f^{-1}(t))\equiv g(t).
\ee
It is easy to check that the function $g$ defined by (\ref{x(t)}) is left invariant if one reparameterizes the world line: $l=\{C'(\xi ');\,\xi '\in \mathrm{I}' \}$ with $\xi '=\psi (\xi )$, $\psi $ being an invertible smooth function such that  $C'\circ \psi =C$.
In an arbitrary affine spatial chart $\phi $, defined by the origin point $x_0 \in \mathrm{A}^3$ and the linear bijection $L: \mathrm{E}^3\rightarrow \mathbb{R}^3$, we thus have the coordinate vector 
\be\label{space chart}
{\bf x}(t) \equiv \phi (g(t)) = L(g(t)-x_0).
\ee
Then, as in basic textbooks, we can define the velocity vector:
\be\label{velocity chart}
{\bf u}(t) \equiv \frac{\dd {\bf x}}{\dd t}.
\ee
If we change the affine spatial chart by changing the origin point $x_0$ and the linear bijection $L$: $x_0 \hookrightarrow \tilde{x}_0,\ L \hookrightarrow \tilde{L}$, we get easily from (\ref{space chart}):
\be\label{change chart}
\tilde{{\bf x}}={\bf G.x}+{\bf a},\quad {\bf G}\equiv \tilde{L}\circ L^{-1} \in \mathcal{L}(\mathbb{R}^3)\simeq {\sf M}(3,\mathbb{R}), \quad {\bf a}\equiv \tilde{L}(x_0-\tilde{x}_0),
\ee
where the real $3\times 3$ matrix ${\bf G}$ and the point ${\bf a}\in \mathbb{R}^3$ are, of course, independent of the time $t$. It hence follows from (\ref{velocity chart}) and (\ref{change chart}) that 
\be\label{change velocity}
\tilde{{\bf u}}(t) \equiv \frac{\dd \tilde{{\bf x}}}{\dd t} = {\bf G.u}(t),
\ee
or
\be
\tilde{u}^i(t) = \frac{\partial \tilde{x}^i}{\partial x^j} u^j(t) \quad \ (i=1,2,3;\ \mathrm{sum\ on\ }j=1,2,3),
\ee
thus giving an elementary proof that the definition (\ref{velocity chart}) defines indeed a unique vector $u(t)=L^{-1}({\bf u}(t)) = \tilde{L}^{-1}(\tilde{{\bf u}}(t)) \in \mathrm{E}^3$, independently of the particular (affine) chart used. (This can be proved generally by using the formalism of differentiable manifolds.)\\
 
Similarly, the {\it acceleration} of the moving point is defined
as
\be\label{acce}
{\bf a}(t) \equiv \frac{\dd {\bf u}}{\dd t}.
\ee
This also defines a vector $a(t)\in \mathrm{E}^3$ that does not depend on the affine spatial chart, since on changing that chart we get from (\ref{change velocity}):
\be\label{change acce}
\tilde{{\bf a}}(t) \equiv \frac{\dd \tilde{{\bf u}}}{\dd t} = {\bf G.a}(t).
\ee 
Consistently with the mere affine structure, no metric has been used: neither on the spacetime $\mathrm{V_{N-L}}$ or its components $\mathrm{A}^1$ and $\mathrm{A}^3$, nor on the arithmetic spaces $\mathbb{R}^4$, $\mathbb{R}$ and $\mathbb{R}^3$. 
\footnote{\
The derivative of a function $t\mapsto {\bf u}(t)$, from an open interval $\mathrm{I} \subset \mathbb{R}$ into an $\mathbb{R}^n$ space, is defined by the usual limit (when it exists). It hence depends only on the topology on $\mathbb{R}$ and the separated topology on $\mathbb{R}^n$. As is well known, on any $\mathbb{R}^m$ space there is only one separated topology that is compatible with the structure of vector space. Thus the derivatives (\ref{velocity chart}) and (\ref{acce}) do not depend on any particular distance, even less on a metric.
}

  \subsubsection{Moving Spaces in the Affine Spacetime}\label{Moving spaces}

We have to define the ``moving spaces", especially the ones having a translation with uniform and constant velocity \Couleur{$v \in \mathrm{E}^3$}, with respect to the ``absolute space" $\mathrm{A}^3$. This is classically done by defining the trajectories at a constant velocity, hence 
in view of (\ref{x(t)}) it can also be done by using the foregoing definition of spatial points as world lines. The moving space \Couleur{$\mathrm{M}_v$} can be defined as the set of the world lines, each of which represents a trajectory at a constant velocity \Couleur{$v$}:
\be
\mathrm{M}_v\equiv \{l_{x \,v};\, x\in \mathrm{A}^3 \}
\ee
with
\be\label{l_xv}
l_{x \,v} \equiv \{(\theta ^{-1}(t), x+ tv);\, t\in \mathbb{R} \} \subset \mathrm{V}_\mathrm{N-L}.
\ee 
Here, \Couleur{$t\in \mathbb{R}$} is the time and \Couleur{$\theta: \mathrm{A}^1\rightarrow \mathbb{R}$} is a given ``time chart": see around Eq. (\ref{Change time}). 
\footnote{\
If one changes the time chart by (\ref{Change time}), the definition (\ref{l_xv}) applied with $\theta '$ in the place of $\theta $ gives $l'_{x \,v} =l_{x+bv \ av}$, as one checks easily. Therefore, the set $\mathrm{M}'_v$ of the lines $l'_{x \,v}$ is just $\mathrm{M}_{v'}$ with $v'\equiv av$.
}
Note that, from (\ref{space chart}) and (\ref{velocity chart}), the velocity of the moving point having the world line $l_{x \,v}$ is represented in the arbitrary affine spatial chart $\phi $ by the constant
\be
{\bf v}(t) = \frac{\dd }{\dd t} \left ( L(x+tv-x_0) \right ) = L(v).
\ee
It corresponds indeed with the constant vector $v\in \mathrm{E}^3$. If \Couleur{$v=0$}, then \Couleur{$l_{x \,v}=l_x$}: the world line defined in (\ref{l_x}). Thus the ``absolute space" or ``ether" is \Couleur{$\mathrm{M}\equiv \mathrm{M}_{v=0}$}, the set of the lines $l_x$. This set is a three-dimensional affine space by the mapping
\be\label{Affine M}
\mathrm{M} \times \mathrm{E}^3 \rightarrow  \mathrm{M}:\quad (l_x,v)\mapsto l_x+v \equiv l_{x+v},
\ee
where $\mathrm{A}^3 \times \mathrm{E}^3 \rightarrow \mathrm{A}^3 : (x,v)\mapsto x+v$ is the mapping which makes $\mathrm{A}^3$ an affine space. It follows from the definition (\ref{Affine M}) that, by mapping $x \in \mathrm{A}^3$ to $F(x)\equiv l_x \in \mathrm{M}$, one defines an affine isomorphism $F$ of $\mathrm{A}^3$ onto $\mathrm{M}$. That isomorphism is ``canonical" in the sense that its definition does not involve an arbitrary choice. Thus, the two representations of space as $\mathrm{A}^3$ or as the set of the world lines $l_x$ are equivalent.\\

From any product chart (\ref{Product chart}) of the spacetime $\mathrm{V_{N-L}}$, we deduce a chart that is relevant to the uniformly moving space $\mathrm{M}_v$, by setting for any event $X=(T,x')\in \mathrm{V_{N-L}}$:
\be\label{Phi_v}
\Phi_v (T,x') \equiv \Phi (T,x'-tv)=(t,\phi (x'-tv)), \quad t \equiv \theta (T).
\ee
The spatial part of the chart is thus:
\be\label{chi_v}
(T,x') \mapsto {\bf x}'\equiv \mathrm{Pr}_2(\Phi_v (T,x'))=\phi (x'-tv).
\ee
By the definition (\ref{l_xv}), this is the constant ${\bf x}'=\phi (x)\in \mathbb{R}^3$, whenever $X=(T,x')$ remains in a given line $l_{x \,v}$. Note that $x$ is the position in the space $\mathrm{A}^3$ of the uniformly moving point defined by the world line $l_{x \,v}$, at time $t\equiv \theta (T)=0$. 
\footnote{\
In the product chart (\ref{Product chart}) the line $l_{x \,v}$ writes:
\be
\Phi (l_{x \,v}) = \{(t,{\bf x}+t{\bf v});\, t\in \mathbb{R}\},\quad {\bf x}\equiv L(x-x_0).
\ee
Hence, the (constant) spacetime vector field $U$ whose components in the chart $\Phi $ are
\be
U^0=1,\quad U^i = v^i
\ee
is a tangent vector field to $l_{x \,v}$, or equivalently $l_{x \,v}$ is a (maximal) integral curve of $U$. Therefore, the constancy on $l_{x \,v}$ of ${\bf x}'$ defined in Eq. (\ref{chi_v}) means exactly that the chart $\Phi _v$ is {\it adapted} to the vector field $U$, in the sense defined by Cattaneo \cite{Cattaneo1969}. See \S \ \ref{global space}\quad below.
}
Thus, given any product affine chart, i.e. having the form (\ref{Product chart}), the definition of the chart (\ref{Phi_v}) adapted to the uniformly moving space involves little arbitrary. (Changing the origin of time is accomplished by changing the time chart $\theta $, see Eq. (\ref{Change time}).) This allows us to define the velocity, with respect to the uniformly moving space $\mathrm{M}_v$, of the particle having the general world line (\ref{World line l}), parameterized by the time $t$ according to (\ref{x(t)}). This is defined using any adapted chart of the form (\ref{Phi_v}) or rather its spatial part (\ref{chi_v}):
\be\label{u'}
{\bf u}'(t) \equiv \frac{\dd {\bf x}'}{\dd t},\quad {\bf x}'(t) \equiv \phi (g(t)-tv)=L(g(t)-tv-x_0).
\ee
One shows immediately, just like in Eqs. (\ref{change chart}) and (\ref{change velocity}), that this defines a unique vector $u'(t)\equiv L^{-1}({\bf u}'(t))\in \mathrm{E}^3$, independently of the affine spatial chart i.e. of the origin point $x_0\in  \mathrm{A}^3$ and the linear bijection $L: \mathrm{E}^3 \rightarrow \mathbb{R}^3$. It also follows immediately from (\ref{u'}) (i) that $u'(t)$ is constantly zero for a moving point ``bound with $\mathrm{M}_v$" (in fact {\it belonging} to $\mathrm{M}_v$), i.e., for a particle following an $l_{x \,v}$ world line; and (ii) that we have for a general particle:
\be\label{u' vs u}
{\bf u}' = {\bf u} - {\bf v}, \quad \mathrm{or}\quad u'=u-v.
\ee 
This, of course, is the addition velocity formula of classical kinematics.\\

\vspace{1mm}
Similarly with (\ref{acce}), we define the acceleration vector $a'(t)$ --- relative, now, to the uniformly moving space $\mathrm{M}_v$ --- from its component vector ${\bf a}'(t) \equiv \frac{\dd {\bf u}'}{\dd t}$, now in any chart having the form (\ref{Phi_v}). We find from  (\ref{u' vs u}) that it is equal to the acceleration $a(t)$ relative to the ``absolute space":
\be
a'(t) = a(t)
\ee
for any moving point, defined by its world line (\ref{World line l}). Thus, the acceleration relative to any moving space \Couleur{$\mathrm{M}_v$} is the same. Therefore, if we postulate Newton's second law with the invariant mass $m$ of classical mechanics: 
\be
{\bf F} = m {\bf a},
\ee
then the invariance of the force ${\bf F}$ follows from that of the acceleration. This is the Galilean relativity in the Newton-Lorentz Universe. In the present formalism, it results from defining the transition to the moving spaces through the charts (\ref{Phi_v}), which involve implicitly the Galilean transformation of the position. Why, then, was Lorentz's first electromagnetic theory incompatible with Galilean relativity? Because Maxwell's equations imply that the velocity of light is a well-defined constant: $c = (\epsilon _0 \mu _0)^{-\frac{1}{2}}$ in SI units --- whereas, with the Galilean transformation, the velocity (of a light ray or a wave front) would depend on the moving space, Eq. (\ref{u' vs u}). Thus, if one uses the Galilean transformation involved in the definition of the charts (\ref{Phi_v}), then Maxwell's equations can be written only in one among the moving spaces $\mathrm{M}_v$.

\subsection{Galileo Universe vs Newton-Lorentz Universe}\label{Galileo}

We just saw that Galilean relativity is well compatible with the ``absolute" product structure (\ref{time-space-product}) for the spacetime: Galilean relativity results from naturally associating with the product spacetime chart (\ref{Product chart}) the spacetime chart (\ref{Phi_v}), that is adapted to the uniformly moving space $\mathrm{M}_v$. (Note that the product spacetime chart (\ref{Product chart}) is hence adapted to the ``fixed" or ``absolute" space $\mathrm{M}$ since $\mathrm{M}\equiv \mathrm{M}_{v=0}$.) At this point, we observe that the privileged status of the ``absolute space" $\mathrm{A}^3 \simeq \mathrm{M}_0$, supposed to result from taking the spacetime as the product (\ref{time-space-product}), is in a sense only apparent: we can start from any among the moving spaces $\mathrm{M}_v$, of course with a given velocity vector $v\in \mathrm{E}^3$, and note that it is a three-dimensional affine space with the same translation space $\mathrm{E}^3$ as the starting ``absolute space" $\mathrm{A}^3$, by the action
\be
\mathrm{M}_v\times \mathrm{E}^3 \rightarrow  \mathrm{M}_v :\quad (l_{x \,v}, u)\mapsto l_{x \,v}+ u \equiv l_{x+u \ v}.
\ee
That affine space, say $\mathrm{A}'^3$, can be used to redefine the Newton-Lorentz Universe as $\mathrm{V}'_\mathrm{N-L}\equiv \mathrm{A}^1\times \mathrm{A}'^3$. Then we can rewrite anything from \S \ref{N-L without R}\quad to \S \ref{Moving spaces}\quad  included, with replacing $\mathrm{A}^3$ by $\mathrm{A}'^3$ and $\mathrm{V}_\mathrm{N-L}$ by $\mathrm{V}'_\mathrm{N-L}$. However, obviously, in this ``second level" definition of the product spacetime, it is now the ``moving space" $\mathrm{M}_v =\mathrm{A}'^3$ with the given vector $v\in \mathrm{E}^3$ which plays the role of the ``absolute space".\\

On the other hand, one can try to make Galilean relativity apparent in the structure of the spacetime itself: instead of defining it as a product, one takes it as a four-dimensional affine space \Couleur{$\mathrm{A}^4$} whose translation space \Couleur{$\mathrm{E}^4$} is endowed with a (relative) time map \Couleur{$ \tau :\mathrm{E}^4\rightarrow \mathbb{R}$} \cite{Arnold1976}. The time interval between two events \Couleur{$X,Y\in \mathrm{A}^4$} is defined to be 
\be
\delta t(X,Y) = \tau(Y-X).
\ee
Since any two affine spaces of equal dimension are isomorphic, \Couleur{$\mathrm{A}^4$} is isomorphic as an affine space to the Newton-Lorentz spacetime \Couleur{$\mathrm{V}_\mathrm{N-L}=\mathrm{A}^1 \times \mathrm{A}^3$}. However, there is no {\it canonical} (preferred) isomorphism of \Couleur{$\mathrm{A}^4$} onto \Couleur{$\mathrm{A}^1\times \mathrm{A}^3$}. Comparing these two spacetime structures in more detail would need to compare the transition from one moving space to another one in the two approaches. Unfortunately, this question is not discussed in Ref. \cite{Arnold1976}. \\

In both the Newton-Lorentz and the Galileo Universe there is additionally a three-dimensional Euclidean metric \Couleur{${\bf h}$}:  

\bi
\item For the Newton-Lorentz Universe, the metric \Couleur{${\bf h}$} acts primarily on the translation space \Couleur{$\mathrm{E}^3$} of the ``absolute space" \Couleur{$\mathrm{A}^3$}.\\

\item For the Galileo Universe, the metric \Couleur{${\bf h}$} acts on the translation space of any of the spaces of simultaneous events, which are three-dimensional affine spaces \cite{Arnold1976}. The same can be done with the \Couleur{${\bf h}$} of the Newton-Lorentz Universe.

\ei
As we saw in Subsect. \ref{N-L}, no metric is needed to define kinematics in the affine spacetime. Of course, the metric is nevertheless needed in classical mechanics, notably to define things as important as are the Euclidean distance, the orthogonal symmetries of the corresponding Euclidean geometry, and the kinetic energy.

\vspace{2mm}
Thus, all in all, the Newton-Lorentz Universe is {\it mathematically less general} than Galileo's, since in Newton-Lorentz but not in Galileo there are canonical ``space'' and ``time" projections. However, in the Newton-Lorentz Universe, Galilean relativity may apply (case of Newton's mechanics) or it may not (case of Lorentz's first electromagnetic theory). Whereas, the Galileo Universe is built to ensure Galilean relativity (no preferred space). A non-Galileo-invariant behaviour may be obtained at the price of enriching the mathematical structure of the Galileo Universe (thus making it less general) --- e.g. by defining preferred time and space projections, thus providing a preferred isomorphism of \Couleur{$\mathrm{A}^4$} onto \Couleur{$\mathrm{A}^1\times \mathrm{A}^3$}. In summary, the Newton-Lorentz Universe is {\it physically more general} and {\it mathematically less general} than is the Galileo Universe. And, physics in the Newton-Lorentz Universe can be either Galileo-relativistic, or not. Hence we may state that {\it at least in classical (non-relativistic) physics, the correspondence between the physics and the mathematical structure of spacetime is not one-to-one.}



\section{Special Relativity}\label{SR}
  \subsection{Minkowski Spacetime}

The Minkowski spacetime is the basic structure of special relativity (SR). The standard view defines it as a four-dimensional vector space endowed with the well-known f\mbox{}lat metric having Lorentzian signature (e.g. \cite{Naber1992}). In the absence of a well-defined ``origin event" in the spacetime of special relativity, it is clearly more correct to define the Minkowski spacetime as the 4-D affine space \Couleur{$\mathrm{A}^4$}, just like the ``Galilean Universe" of Arnold \cite{Arnold1976} --- but now endowed with the f\mbox{}lat (Poincar\'e-)Minkowski metric ${\boldsymbol{\gamma }}$.  For this purpose, \Couleur{$\mathrm{A}^4$} is seen as a differentiable manifold. The defining atlas is made of all 
affine charts (see the definition at \S \ref{Coord_aff}\ \,). As a metric on this manifold, ${\boldsymbol{\gamma}}$ is a mapping associating with any point $X\in \mathrm{A}^4$, a scalar product ${\boldsymbol{\gamma}}_X$ acting on vectors in the tangent space at $X$ to $\mathrm{A}^4$, $\mathrm{T}_X \mathrm{A}^4$. 
\footnote{\ \label{E iso to TV}
In fact, the tangent space $\mathrm{T}_X \mathrm{A}$ to an affine space $\mathrm{A}$ is naturally identified with the unique translation space $\mathrm{E}$ of $\mathrm{A}$, independently of the point $X\in \mathrm{A}$. To see this, remind that $\mathrm{T}_X \mathrm{A}$ is defined as the set of the equivalence classes of the curves $\xi \mapsto C(\xi )\in  \mathrm{A}$ that are defined in an open interval containing $0$ and are such that $C(0)=X$, modulo the relation ``\,$C\sim C'$ if $C$ and $C'$ are tangent at $\xi =0$\," (e.g. \cite{DieudonneTome3}). By associating with any vector $U\in \mathrm{E}$ the equivalence class $\dot{C}$ of the curve $C : \xi \mapsto X+\xi U$, one defines a canonical isomorphism from the translation space \Couleur{$\mathrm{E}$} onto $\mathrm{T}_X \mathrm{A}$. Therefore, ${\boldsymbol{\gamma}}_X$ can be seen as acting on the translation space \Couleur{$\mathrm{E}^4$} of \Couleur{$\mathrm{A}^4$}, independently of $X\in \mathrm{A}^4$.
}\\

Specifically, the Minkowski metric ${\boldsymbol{\gamma}}$ on $\mathrm{A}^4$ can be defined by choosing (arbitrarily) one affine chart $\chi $ of $\mathrm{A}^4$ and by imposing that, at any $X\in \mathrm{A}^4$, the components of \Couleur{${\boldsymbol{\gamma}}_X$} in the chart $\chi$ verify $(\gamma_{\mu\nu}) = {\boldsymbol{\eta}}\equiv \mathrm{diag}(1,-1,-1,-1)$. A {\it Cartesian chart\,} $\chi'$ on \Couleur{$(\mathrm{A}^4,{\boldsymbol{\gamma}})$} is defined to be an affine chart of $\mathrm{A}^4$ such that the metric ${\boldsymbol{\gamma}}$ has that same form. And, the Lorentz transformations are just the changes from one Cartesian chart to another one having the same origin. Now an origin-preserving change of the affine chart is any linear bijection of $\mathbb{R}^4$ onto itself, thus generally is {\it not} a Lorentz transformation. It follows that there are many different Minkowski metrics on the affine spacetime $\mathrm{A}^4$. (The same is true if one defines the Minkowski spacetime as a vector space $\mathrm{E}^4$ instead, thus hiding the choice of the origin: any basis of the vector space can be chosen as one in which the metric's matrix is ${\boldsymbol{\eta}}$.) This remark does not imply a consequential mathematical ambiguity, for the Minkowski space is thus defined up to an isometric affine transformation. Nor does it imply a physical ambiguity, because the metric notions are made  real through clocks and light rays.\\

Note that one may define the {\it space associated with a Cartesian chart $\chi$,} \Couleur{$\mathrm{M}_\chi$},  as a set of world lines, again:
\be\label{M_chi}
\Couleur{\mathrm{M}_\chi\equiv \{l_{\chi\,{\bf x}};\ {\bf x}\in \mathbb{R}^3 \}\quad \mathrm{with}\ l_{\chi\,{\bf x}}=\chi^{-1}(\mathbb{R} \times \{{\bf x}\}) \quad ({\bf x}\in \mathbb{R}^3)}.
\ee 
Thus \Couleur{${\bf x}\in \mathbb{R}^3$} is the common spatial position vector, in the chart \Couleur{$\chi$}, of all events \Couleur{$X\in l_{\chi{\bf x} }$}.\\


  \subsection{Newton-Lorentz Spacetime for Lorentz-Poincar\'e SR}\label{L-P SR}
 Instead of starting from the ``block" affine spacetime $\mathrm{A}^4$, we may also define the Minkowski spacetime from the Newton-Lorentz Universe $\mathrm{V}_\mathrm{N-L}=\mathrm{A}^1\times \mathrm{A}^3$, just like we did for classical kinematics in Subsect. \ref{N-L}. To define the Minkowski metric on that product spacetime, we endow the respective translation spaces \Couleur{$\mathrm{E}^1$} and \Couleur{$\mathrm{E}^3$} with Euclidean metrics, say \Couleur{${\boldsymbol{h}}^1$} on $\mathrm{E}^1$ and \Couleur{${\boldsymbol{h}}^3$} on $\mathrm{E}^3$. (As shown in Note \ref{E iso to TV}, we can identify with $\mathrm{E}^3$ the tangent space to $\mathrm{A}^3$ at any point $x\in \mathrm{A}^3$, and similarly for $\mathrm{E}^1$ and $\mathrm{A}^1$.) Then we define the Minkowski metric on the translation space of \Couleur{$\mathrm{A}^1\times \mathrm{A}^3$}, that is $\mathrm{E}=\mathrm{E}^1\oplus \mathrm{E}^3$. This is a 4-D vector space that, as a set, is the Cartesian product $\mathrm{E}=\mathrm{E}^1\times \mathrm{E}^3$. Any pair of vectors $U,U'$ in $\mathrm{E}$ has the form $\,U=(\tau ,v),\ U'=(\tau ',v')$ with $\tau ,\tau '\in \mathrm{E}^1$ and $v, v' \in \mathrm{E}^3$ and we set
\be\label{Product metric}
{\boldsymbol{\gamma}}(U,U')={\boldsymbol{\gamma}} ((\tau ,v),(\tau ',v'))\equiv {\boldsymbol{h}}^1(\tau ,\tau ')-{\boldsymbol{h}}^3(v,v').
\ee
Note that (\ref{Product metric}) defines ${\boldsymbol{\gamma}}(U,U')$ independently of any basis on  $\mathrm{E}=\mathrm{E}^1\times \mathrm{E}^3$. Naturally, SR and the study of Lorentz transformations lead one to consider, on that vector space, bases whose vectors belong neither to $\mathrm{E}^1 \times \{0_{\mathrm{E}^3}\}$ nor to $\{0_{\mathrm{E}^1}\}\times \mathrm{E}^3$.\\

The {\it Lorentz-Poincar\'e (L-P) version of SR} is a physical theory that starts from the ``ether", seen as an inertial frame $\mathcal{E}$ such that (i) Maxwell's equations are valid in $\mathcal{E}$ (in particular, light propagates at a constant velocity $c$ with respect to $\mathcal{E}$) and (ii) any material object that moves with respect to $\mathcal{E}$ undergoes a Lorentz contraction. One derives first the Lorentz transformation from these assumptions, and then the whole of special relativity follows. See in particular Prokhovnik \cite{Prokhovnik1967} and references therein. A summary of L-P SR is given in Ref. \cite{A9}, Sect. 2. In L-P SR, an important physical concept is that of a ``light clock", which measures the time that it takes for light to go forth and back between two mirrors --- this is basically what an interferometer does. (Light clocks can also be used to justify the ``clock hypothesis" which states that a clock measures the proper time along a trajectory, and which is used not only in SR but also in relativistic gravity \cite{Fletcher2013}.) Once the ``relativistic" effects of SR can thus be derived from ``absolute" metrical effects of the motion through an ``ether", it suggests itself to search for an L-P {\it type} theory of gravitation, in which gravitation also has metrical effects \cite{A9, Broekaert2007}. The  foregoing introduction of the Minkowski spacetime as the product spacetime $\mathrm{A}^1\times \mathrm{A}^3$ endowed with the metric (\ref{Product metric}) is a way to formalize the L-P version of special relativity, as well as to define the background spacetime for L-P type theories of gravitation.\\

\vspace{2mm}
{\it Thus either Galilean relativity or special relativity can be implemented on a common preexisting structure with preferred time and preferred space: 
the affine manifold} \Couleur{$\mathrm{V}_\mathrm{N-L} \equiv \mathrm{A}^1\times \mathrm{A}^3$}.


\section{Curved Spacetime}\label{Curved}
\subsection{Two Dynamics in a General Lorentzian Spacetime}\label{TwoDynamics}

A Lorentzian spacetime is the basic structure in relativistic theories of gravitation. In that case, the spacetime is a general 4-D differentiable manifold \Couleur{$\mathrm{V}$}, endowed with a pseudo-Riemannian metric \Couleur{${\boldsymbol{g}}$} having Lorentzian signature, i.e., $(+ - - -)$ or $(- + + +)$. 
Thus \Couleur{$\mathrm{V}$} is not in general diffeomorphic to any product manifold, and is much less often an affine space. With the mere data \Couleur{$(\mathrm{V},{\boldsymbol{g}})$}, we may define Einstein's dynamics, which consists in assuming geodesic motion for test particles and in assuming, for a continuous medium or a system of fields, the generally-covariant equation
\be
T^{\mu\nu}_{;\nu}=0.
\ee
(Here, $T^{\mu\nu}\ \, (\mu ,\nu =0,...,3)$ are the components of the energy-momentum(-stress) tensor. Semicolon denotes covariant derivative defined with the metric connection associated with ${\boldsymbol{g}}$, which is explicitly given in terms of the Christoffel symbols.)\\

\vspace{3mm} 
However, we may also define another dynamics. In Ref. \cite{A16} an extension of the special-relativistic form of Newton's second law to a general Lorentzian spacetime was defined. This extension can be defined in any admissible reference f\mbox{}luid $\mathcal{F}$, physically defined by Cattaneo \cite{Cattaneo1958} as a 3-D congruence of moving points (fictitious ``observers") having time-like world lines. The extension involves the spatial metric ${\boldsymbol{h}}$ defined in a given reference f\mbox{}luid  $\mathcal{F}$ from the spacetime metric ${\boldsymbol{g}}$. (The definition of ${\boldsymbol{h}}$ is explained e.g. by M\o ller \cite{Moller1952} and by Landau \& Lifshitz \cite{L&L}.) It involves also the definition of a gravity acceleration, which is a spatial vector field ${\bf g}$ in the given reference f\mbox{}luid $\mathcal{F}$. One way to determine the field ${\bf g}$ is to ask that geodesic motion be recovered \cite{A16}. Thus in that case, Einstein's generally-covariant dynamics is recovered and no additional structure is needed, insofar as the reference f\mbox{}luid is arbitrary. But there is a different possible definition of the field ${\bf g}$, in which its form is suggested by an interpretation of gravity as a pressure force in a perfectly f\mbox{}luid ``ether" \cite{A9}. The same form is found if one asks that (i) ${\bf g}$ should not depend on the time variation of the metric ${\boldsymbol{g}}$ and should be linear with respect to the space variation of ${\boldsymbol{g}}$, and (ii) the extension of Newton's second law with this field ${\bf g}$ should imply geodesic motion in the particular case of a static metric \cite{A16}. However, this other definition of the field ${\bf g}$ is covariant only under the coordinate transformations having the form \cite{A16}
\be\label{spatial change +f(t)}
x'^0 = \theta (x^0), \quad x'^i= \psi ^i(x^1,x^2,x^3)\ \, (i=1,2,3).
\ee 
These transformations leave the reference f\mbox{}luid unchanged. Hence, the alternative dynamics based on the second form of the gravity acceleration vector field ${\bf g}$ needs that one assumes a preferred reference f\mbox{}luid. However, the Lorentzian spacetime $(\mathrm{V},{\boldsymbol{g}})$ itself is left unchanged. 


\subsection{Defining the Space in a General Spacetime}\label{Space}

We shall now discuss how to define the 3-D {\it space manifold} associated with a given reference f\mbox{}luid. It is very useful for relativistic theories of gravitation, and in our opinion is even necessary for them --- e.g. when one wants to define the space of states for quantum theory in a curved spacetime \cite{A44,A52}. This is  also necessary, among other things, to define precisely and naturally the notion of spatial tensor: a spatial tensor is for us simply a tensor on the space manifold \cite{A16}. 
For example, such spatial tensors are used in the alternative dynamics just mentioned --- e.g. the spatial metric ${\boldsymbol{h}}$ (also used in more standard relativistic theories of gravitation) or the gravity acceleration vector field ${\bf g}$. Without using the notion of the space manifold associated with a reference f\mbox{}luid, another notion of spatial tensor can be defined, namely as a {\it spacetime} tensor which is equal to its spatial projection \cite{Cattaneo1958}. (The projection tensor has been defined by Cattaneo \cite{Cattaneo1958}, it too depends on the reference f\mbox{}luid.) 
\footnote{\
Objects with merely spatial indices, such as the ``spatial metric tensor" and the ``3-velocity vector", had been considered in the literature on relativistic gravitation before Cattaneo \cite{Cattaneo1958}, e.g. by Landau \& Lifshitz \cite{L&L} and by M\o ller \cite{Moller1952}. However, the sense in which these objects could really be {\it tensors} was not and could not be defined before having a precise notion of spatial tensor --- be it in the sense defined by Cattaneo \cite{Cattaneo1958} or as a tensor on the space manifold. It was noted by Cattaneo \cite{Cattaneo1958}, but apparently not by Landau \& Lifshitz \cite{L&L} and by M\o ller \cite{Moller1952}, that these objects indeed transform tensorially on a change of spatial coordinates which does not depend on the time coordinate. 
}
But this other concept (used also after Cattaneo \cite{Cattaneo1958}, e.g.  \cite{Massa1974a, Massa1974b, Mitskievich1996, JantzenCariniBini1992}) is far less simple, especially when it comes to the definition of a relevant covariant derivative of spatial tensors. See the discussion in Ref. \cite{A16}, Sect. 4.  \\

A case where the space manifold is obvious to define is got \cite{A15} when one assumes that the spacetime manifold is the product $\mathrm{V} = \mathbb{R} \times \mathrm{M}$, with $\mathrm{M}$ a three-dimensional differentiable manifold. 
\footnote{\
This turns out to be the case, in particular, when $\mathrm{V} $ is endowed with a Lorentzian metric ${\boldsymbol{g}}$ such that the Lorentzian manifold $(\mathrm{V}, {\boldsymbol{g}})$ be {\it globally hyperbolic}. More precisely, it has been proved by Bernal and S\'anchez \cite{BernalSanchez2003} that any connected Lorentzian manifold that is globally hyperbolic is diffeomorphic to such a product $\mathbb{R} \times \mathrm{M}$. (The ``topological" case, with the diffeomorphism being replaced by a mere homeomorphism, had been treated by Geroch \cite{Geroch1970}.)
}
Similarly, in Subsect. \ref{N-L}, a framework for classical kinematics was defined as the product of the two affine spaces $\mathrm{V}=\mathrm{A}^1\times \mathrm{A}^3$. It was noted in Subsect. \ref{L-P SR} that this is also the natural framework for Lorentz-Poincar\'e SR. 
In the present more general situation, by defining $\mathrm{V} = \mathrm{A}^1 \times \mathrm{M}$ instead of $\mathrm{V} = \mathbb{R} \times \mathrm{M}$, one would avoid fixing an origin of time and a time unit; cf. Eq. (\ref{Change time}). 
The manifold $\mathrm{M}$ defines a preferred reference f\mbox{}luid, whose world lines are the lines $\mathbb{R}\times \{x\}$ (or $\mathrm{A}^1 \times \{x\}$) for some $x\in \mathrm{M}$, just like the lines (\ref{l_x}) in the case of the affine space. \\

However, this way of definition is restricted to the case of a product spacetime, and even in that case it is restricted to the ``unmoving" space $\mathrm{M}$. To implement this definition in a general reference f\mbox{}luid in a general Lorentzian spacetime needs a definition of what is understood by ``space" in such a general context. Essentially: as described in the Introduction, a three-dimensional congruence of world lines provides a reference f\mbox{}luid according to Cattaneo's view \cite{Cattaneo1958}. We define the associated space manifold, say $\mathrm{N}$, as the very set of these world lines; we endow the spacetime with a set of charts, for any of which the spatial coordinates remain constant on any among those world lines; we ensure that the spatial part of such a chart is then a chart on $\mathrm{N}$ \cite{A44,A52}. Thus the space manifold $\mathrm{N}$ is browsed by precisely the triplet ${\bf x}\equiv (x^j)$ made with the spatial projection of the spacetime coordinates. The spatial tensors are defined simply as tensor fields on the space manifold $\mathrm{N}$.

  \subsubsection{Defining a Local Space in a General Spacetime}\label{local space}

In a local approach, we may start from a coordinate system (as was also the case with Cattaneo's approach), i.e., from a chart \Couleur{$\chi : X \mapsto {\bf X}=(x^\mu)\in\mathbb{R}^4$}, whose domain $\mathrm{U}$ is an open subset of the spacetime manifold $\mathrm{V}$. No metric is necessary. As in Eq. (\ref{M_chi}), consider the world lines:
\be\label{l_chi x}
l_{\chi \,{\bf x}}\equiv \chi^{-1}(\mathbb{R} \times \{{\bf x}\}).
\ee
Thus again, $l_{\chi \,{\bf x}}$ is the curve of the spacetime, made by the events for which the vector  of the spatial coordinates (in the chart $\chi $) is a given triplet ${\bf x}\in \mathbb{R}^3$. The difference with (\ref{M_chi})$_2$ is that now $\chi $ is a fully general chart on a fully general manifold $\mathrm{V}$, in particular its domain $\mathrm{U}$ is in general only an open subset of $\mathrm{V}$. One shows easily that $l_{\chi \,{\bf x}}$ is indeed a world line in the sense that it is the image of an open subset of the real line $\mathbb{ R}$ by a smooth mapping \cite{A44}. But, in general, it is not a connected set, i.e., the curve (\ref{l_chi x}) may well consist of several pieces. However, any connected component of $l_{\chi \,{\bf x}}$ is an integral curve of the vector field $U$ having components $U^\mu =\delta ^\mu _0$ in the chart $\chi $. Thus the set of the world lines (\ref{l_chi x}) is a reference f\mbox{}luid: its particles are the ones whose trajectory is fixed at some position ${\bf x}$ in the chart $\chi $.\\

Consider the charts which have the same domain $\mathrm{U}$ as the starting chart $\chi $ and which exchange with $\chi $ by a purely spatial coordinate change, i.e., Eq. (\ref{spatial change +f(t)})$_2$\, plus \,$x'^0=x^0$. These charts build a set $\mathrm{F}$, which is an equivalence class for the relation ``$\chi' \sim \chi ''$ iff $\chi' $ and $\chi ''$ have domain $\mathrm{U}$ and exchange by a purely spatial coordinate change". We call such an equivalence class  a {\it reference frame} $\mathrm{F}$, since indeed this definition corresponds with the notion of reference frame that is relevant to quantum mechanics in a curved spacetime, i.e., a reference f\mbox{}luid endowed with a given time map \cite{A44}.  It is easy to show that the set of the lines (\ref{l_chi x}) is unchanged if one substitutes for $\chi $ any chart $\chi '$ belonging to the set $\mathrm{F}$. We thus define the set 
\be 
\mathrm{M}_\mathrm{F} \equiv \{l_{\chi \,{\bf x}};\ {\bf x}\in \mathbb{R}^3\} = \{l_{\chi' \,{\bf x}'};\ {\bf x}'\in \mathbb{R}^3\} \quad (\chi '\in \mathrm{F}).
\ee

\vspace{2mm}
We may define a natural structure of three-dimensional differentiable manifold on \Couleur{$\mathrm{M}_\mathrm{F}$}, for which the defining atlas is made of the mappings $\widetilde{\chi'}: l_{\chi' \,{\bf x}'} \mapsto {\bf x}'$, for any $\chi '\in \mathrm{F}$ \cite{A44}.



  \subsubsection{Defining a Global Space in a General Spacetime}\label{global space}

Consider now a global reference f\mbox{}luid, defined by a  global, {\it non-vanishing} vector field \Couleur{$U$} on the spacetime manifold $\mathrm{V}$. Again, no metric
is necessary. The maximal integral curves \Couleur{$l$} of the vector field $U$ are those integral curves of $U$ for which the interval of definition $\mathrm{I}$ is the largest possible, also called the {\it orbits} of $U$. (Once more, the relevant curve is the {\it image} $l\equiv C(\mathrm{I})$ with $C$ verifying $\frac{\dd C}{\dd \xi }=U(C(\xi ))$ for $\xi \in \mathrm{I}$.) The global space manifold associated with \Couleur{$U$} is just the set \Couleur{$\mathrm{N}_U$} of these curves \cite{A52}. Call a chart \Couleur{$\chi$} on \Couleur{$\mathrm{V}$} ``$U$-adapted" iff for any curve \Couleur{$l\in\mathrm{N}_U$} the spatial coordinates vector \Couleur{${\bf x}$} is constant on \Couleur{$l$}. In that case, one hence deduces from the chart $\chi$, in a unique way, a map 
\be
\bar{\chi}: \mathrm{N}_U \rightarrow \mathbb{R}^3,\ l \mapsto {\bf x}.
\ee
Thus in the same way as for a Cartesian chart [see after Eq. (\ref{M_chi})], ${\bf x}\in \mathbb{R}^3$ is the common spatial position vector, in the $U$-adapted chart \Couleur{$\chi$}, of all events $X\in l$. If an $U$-adapted chart $\chi $ is such that $\bar{\chi}$ is injective, so that a data ${\bf x}$ (taken in the image $\bar{\chi}(\mathrm{N}_U)$) determines just one curve \Couleur{$l\in\mathrm{N}_U$}, we say that $\chi $ is ``nice". The set of the nice $U$-adapted charts is denoted by $\mathcal{F}_U$.\\

By investigating the conditions under which there exists a nice $U$-adapted chart in the neighborhood of an arbitrary event $X\in \mathrm{V}$, one is led to define a concept of {\it ``normal"} non-vanishing vector field \Couleur{$U$}. Essentially, $U$ is normal if any event $X\in \mathrm{V}$ has a neighborhood $\mathrm{U}$ such that the intersection with $\mathrm{U}$ of any orbit of $U$ be connected. (The precise definition is given in Ref. \cite{A52}.) If $U$ is normal, it can be proved that the maps $\bar{\chi}$, each of which is deduced from a nice $U$-adapted chart $\chi $ \,($\chi \in \mathcal{F}_U$), build an {\it atlas} of \Couleur{$\mathrm{N}_U$} 
\cite{A52}. \\

An important example of normal vector field is the following one \cite{A52}. (i) Assume that there is a global chart $\chi $ of the spacetime manifold $\mathrm{V}$. Then, the global vector field $U$ that has components $U^\mu =\delta ^\mu _0$ in the chart $\chi $, which is tangent to the lines (\ref{l_chi x}), is a normal vector field. Further, suppose (ii) that each among the lines (\ref{l_chi x}) is {\it connected}.  Then $\chi $ is a nice $U$-adapted chart, moreover the set $\mathrm{N}_U$ endowed with the charts $\,\overline{\chi'} \ \, (\chi '\in \mathcal{F}_U)\,$ is a (3-D) differentiable manifold. This result may seem unsurprising in view of the results summarized in \S \ref{local space}\  \,, but it is not an application of those, as is apparent from the necessity of assuming the connectedness of the lines (\ref{l_chi x}). Moreover it covers a number of relevant situations in relativistic theories of gravitation. It applies even to some singular spacetimes of general relativity, e.g. it applies to the ``maximally extended" Schwarzschild manifold endowed with the Kruskal-Szekeres coordinates \cite{A52}. Due to a plausibility argument based on some proved theorems of ``transversality"  \cite{A52}, we expect that much more general normal vector fields do exist.\\

The relation between the local space manifold $\mathrm{M}_\mathrm{F}$ associated with a reference frame $\mathrm{F}$ (in the precise sense defined in \S \ref{local space}\ \,) and the global space manifold $\mathrm{N}_U$ associated with a normal vector field $U$ has been studied.  If the charts \Couleur{$\chi$} making the class $\mathrm{F}$ are nice \Couleur{$U$}-adapted charts (which means that the local reference frame $\mathrm{F}$ and the global vector field $U$ correspond with the same three-dimensional network of reference points), then the local space manifold \Couleur{$\mathrm{M}_\mathrm{F}$} is an open subset of the global space manifold \Couleur{$\mathrm{N}_U$} \cite{A52}. Moreover, each world line in $\mathrm{M}_\mathrm{F}$ is the intersection with the local domain $\mathrm{U}$ of a world line in $\mathrm{N}_U$, i.e., of an orbit of the global vector field $U$. (Recall that here $\mathrm{U}$, an open subset of the spacetime, is the common domain of all charts belonging to the local reference frame $\mathrm{F}$.) Thus the relation between the two concepts is as good as it might be hoped.



\section{Conclusion}

The ``Newton-Lorentz Universe", defined as the product $\mathrm{A}^1\times \mathrm{A}^3$ of two affine spaces, provides a simple and rigorous framework to define classical (non-relativistic) kinematics. No metric is needed even to define the acceleration. The Galileo invariance comes up with the natural choice (\ref{Phi_v}) for an affine coordinate system adapted to a uniformly moving frame. However, that same affine space $\mathrm{A}^1\times \mathrm{A}^3$ can be endowed with the Minkowski metric (\ref{Product metric}), thus becoming the arena for special relativity as well. More generally: although spacetime is a very useful and clever mathematical concept, the geometric structure of spacetime as this is usually understood (thus including the data of a metric) is not uniquely related with the physics, especially regarding the validity of the relativity principle. On the same spacetime (the Newton-Lorentz Universe, the translation space \Couleur{$\mathrm{E}^3$} of the ``absolute space" \Couleur{$\mathrm{A}^3$} being endowed with an Euclidean metric), physics may be either (Galileo-)relativistic or not. Conversely, the same physics can be implemented on different spacetime structures, e.g. Galileo-invariant mechanics can be implemented on the product affine spacetime $\mathrm{A}^1\times \mathrm{A}^3$ as well as on the ``block" affine spacetime $\mathrm{A}^4$. Moreover, just the same is true for special-relativistic physics, i.e., it too can be implemented on $\mathrm{A}^1\times \mathrm{A}^3$ as well as on $\mathrm{A}^4$. (Of course the metric differs between Galilean and special relativity, as is apparent in the above discussion.) On a Lorentzian spacetime, a preferred-frame dynamics can be defined, as an alternative to the Einstein dynamics. \\

In physics we need a definite notion of {\it space} \cite{A44,A52}. In the absence of a proper definition of that in relativistic physics, practitioners use the triplet of the spatial coordinates, \Couleur{${\bf x}\equiv (x^i)$}. It should be with guilty conscience because, the coordinate system being arbitrary (especially in relativistic theories), this {\it a priori} does not designate any coordinate-independent object. However, it can actually be justified in the proposed framework. The general definition of space is got from a 3-D congruence of reference trajectories, that constitutes a {\it reference f\mbox{}luid} \cite{Cattaneo1958}. This is directly ``physical", because the trajectories can be concretely realized. The relation with the spacetime concept is that the space associated with a reference f\mbox{}luid is formally defined as the set of the reference world lines \cite{Norton1985, A16}, and that set can be endowed with a natural structure of differentiable manifold \cite{A44, A52}. (There are both a local and a global definition of the associated space, but the two are nicely related together \cite{A52}.) This definition of physical space implies of course that it depends on the reference f\mbox{}luid. In {\it adapted} coordinate systems, each point of this space is specified by the triplet \Couleur{${\bf x}$}, thus justifying the definition commonly used in practice. Thus, once a reference f\mbox{}luid is given, we may define a spatial differential geometry on that manifold browsed by \Couleur{${\bf x}$}. In particular we may define spatial tensors simply as tensor fields on the associated space manifold.



\end{document}